\documentclass[a4paper,11pt]{article}
\usepackage{jheppub}
\usepackage{eurosym}
\usepackage{amssymb}
\usepackage{amsfonts,cite}
\usepackage{amsmath}
\usepackage{graphicx}
\usepackage{multirow}
\usepackage{xcolor}
\usepackage{epsfig}
\usepackage{fancyhdr}

\DeclareMathOperator\sech{sech}
\DeclareMathOperator\csch{csch}

\newbox\mybox

\newcommand\fverb{\setbox\mybox=\hbox\bgroup\verb}
\newcommand\fverbdo{\egroup\medskip\noindent\fbox{\unhbox\mybox}\ }
\newcommand\fverbit{\egroup\item[\fbox{\unhbox\mybox}]}
\abstract{We present a novel approach for constructing quasi-isospectral higher-order Hamiltonians from time-independent Lax pairs by reversing the conventional interpretation of the Lax pair operators. Instead of treating the typically second-order $L$-operator as the Hamiltonian, we take the higher-order $M$-operator as the starting point and construct a sequence of quasi-isospectral operators via intertwining techniques. This procedure yields a variety of new higher-order Hamiltonians that are isospectral to each other, except for at least one state. We illustrate the approach with explicit examples derived from the KdV equation and its extensions, discussing the properties of the resulting operators based on rational, hyperbolic, and elliptic function solutions. In some cases, we present infinite sequences of quasi-isospectral Hamiltonians, which we generalise to shape-invariant differential operators capable of generating such sequences. Our framework provides a systematic mechanism for generating new integrable systems from known Lax pairs.}    

\title{\bfseries Quasi-isospectral higher-order Hamiltonians via a reversed Lax pair construction}

\author[a]{Francisco Correa,}
\author[b]{Andreas Fring}

\affiliation[a]{Departamento de F\'isica,   Universidad de Santiago de Chile, Av. Victor Jara 3493, Santiago, Chile}
\affiliation[b]{Department of Mathematics, City St George's, University of London, Northampton Square, \\ London EC1V 0HB, UK}

\emailAdd{hipertuga@gmail.com}
\emailAdd{a.fring@city.ac.uk}

\begin{document}
		\maketitle
	
	\pagestyle{fancy}
	\fancyhead{} 
	\fancyhead[LE,RO]{\small\itshape  Quasi-isospectral higher-order Hamiltonians via a reversed Lax pair construction} 
	
	\renewcommand{\headrulewidth}{0.4pt}

\section{Introduction}

The exploration of higher-order Hamiltonians has received renewed interest, particularly in the context of integrable models \cite{bethanAF,fring23int}. These systems exhibit interesting properties in their own right, but the studies gain additional traction when one makes use of the idea to exchange space and time \cite{samoil1991,avan2016lagrangian,smilga2021exactly,damour2022dyn,fring2024higher,fring2024nonlinear}, as in that setting they constitute a large class of higher time-derivative theories, e.g. \cite{smilga2017classical}, which in turn have potential relevance to fundamental physics, including quantum gravity. 
 
In the conventional approach one usually takes the higher charges as candidates for higher order Hamiltonians. Recalling that the traditional approaches to integrable systems often revolve around the Lax pair formalism, wherein the second-order Lax operator $L$ is typically identified as the Hamiltonian. In this work, we present a novel and reversed perspective: we treat the higher-order Lax operator $M$ as the primary Hamiltonian and build a hierarchy of quasi-isospectral systems through intertwining techniques.  Some physical properties of the higher-order Lax operators $M$ as conserved charges, which arise naturally from the Lax equation in time-independent hierarchies, has been explored in various contexts. These operators encode spectral information in a simpler and comprehensive manner compared to the standard second-order Hamiltonians. Features such as degeneracies in reflectionless and finite-gap systems \cite{correa1, correa2, correa3}, as well as resonances and spectral singularities \cite{correa4,correa5}, can be captured by these higher-order charges. This richer structure provides strong motivation to consider them as Hamiltonians in their own right. The existence of these conserved charges can be understood from the fact that two quantum Hamiltonians may be connected by more than one type of intertwining operator with different differential orders \cite{correa2,correa3,correa4}. This is not a surprise since it is also known that Hamiltonians can be factorized in terms of different intertwining operators, see also \cite{andrianov2003nonlinear,ioffe2004susy,acar2023unusual}.

Our approach here differs from well explored ideas of taking the intertwining operators to be of higher order, but keeping the Hamiltonian in its standard second order form, see for instance  \cite{correa1,correa2,correa3,correa4,correa5,andrianov2003nonlinear,ioffe2004susy,acar2023unusual,andrianov1997matrix,carinena2007isochronous,roychoudhury2008pseudo,bagchi2009existence,inzunza2021conformal,hussin2022third}.

This reversed Lax pair construction enables the systematic derivation of new families of quasi-isospectral higher-order Hamiltonians, which differ from each other by at least one state, commonly the ground state. We demonstrate the method by applying it to the time-independent Korteweg-de Vries (KdV) equation and its extensions, employing rational, hyperbolic, and elliptic solutions. Our approach also allows for the generation of infinite sequences of such Hamiltonians and provides a generalized framework based on shape-invariant differential operators, offering new avenues for constructing integrable systems beyond the traditional Lax pair paradigm.

\section{From Lax pairs to quasi-isospectral higher-order Hamiltonians}
The central property of a classical integrable system is the existence of a Lax pair \cite{Lax}, as it facilitates the derivation of all conserved quantities of the model. By definition a Lax pair requires the existence of two operators $L$ and $M$ satisfying the equations 
\begin{equation}
	\dot{L} = \left[ M ,L \right], \qquad \Leftrightarrow  \qquad L \phi = \lambda \phi , \quad\dot{\phi} = M \phi,   \label{Lax}
\end{equation} 
 up to the validity of the respective equations of motion. Overdots denote time-derivatives, $\phi$ is an auxiliary function and crucially the eigenvalues $\lambda$ are always time-independent so that the first equation on the right hand side can be interpreted as an instantaneous eigenvalue equation. In this sense and further augmented by the fact that $L$ is a second order differential operator in coordinate space involving a function that can be interpreted as potential, $L$ is the natural operator to be viewed as  Hamiltonian. Moreover, all the quantities $Q_k := \text{tr}(L^k) $ are conserved, $\dot{Q}_k=0$, and in involution $\{Q_i,Q_j \}=0$, manifesting the integrable nature of the system. For more detail on integrable systems see e.g. \cite{OP2}.

 Here we consider the time-independent scenario and interpret at first the Lax operator $L \equiv H$ as a Hamiltonian so that in this case (\ref{Lax}) reduces to 
 \begin{equation}
 	\left[ H ,M \right]=0, \qquad \Leftrightarrow  \qquad L \phi = \lambda \phi , \quad H \left( M \phi\right) = \lambda   \left( M \phi\right) .   \label{Laxti}
 \end{equation} 
Even though $M$ is in general of higher order in $x$-derivatives, subsequently we also interpret it as a Hamiltonian and construct a {\em quasi-isospectral} operator $\tilde{M}$ for it.  Quasi-isospectral means here that we loose at least one state in the procedure, that is typically the ground state. Finally, we construct a new Hamiltonian that commutes with $\tilde{M}$ by means of intertwining techniques. We will now outline the procedure. The starting point is always the time-independent Lax equation
\begin{equation}
	\left[ L ,M \right]=0, \label{LMcom}
\end{equation} 
and we will address the question of how to construct the relevant intertwining structure with reversed roles of $L$ and $M$.
\subsection{The standard intertwining operator approach}
In order to highlight the difference between the well-known standard intertwining operator approach and our altered version, we start by reviewing the conventional approach. When deconstructed to its bare minimum it consists of the following four principal steps:
\begin{itemize}
  \item[S1:] Find an operator $L_+$ with common ground state $\psi_0$ to $L$ 
     \begin{equation}
     	L \psi_0      =0,  \qquad     	L_+ \psi_0      =0.
     \end{equation} 
    \item[S2:] Use $L_+$ as a left intertwining operator and find a new operator $\tilde{L}$ so that
    \begin{equation}
    	L_+ L       =\tilde{L} L_+.
    \end{equation}  
      \item[S3:] Find a second right intertwining operator $L_-$ so that
    \begin{equation}
    L  L_- = 	L_- \tilde{L}    .
    \end{equation}  
     \item[S4:] Identify a new operator $\tilde{M}$ that commutes with $\tilde{L}$
      \begin{equation}
     \left[ \tilde{L} ,\tilde{M} \right]=0.
     \end{equation} 
     \item[S5:] Repeat the steps S1-S4 by starting with $\tilde{L}$ instead of $L$ to find a new operator $\hat{L}$.
\end{itemize}
There are two formal solutions to the above scheme:
\indent     \begin{itemize}
	\item[Sol1:] 
	\begin{equation}
		\tilde{M}       = L_+  L_-, \qquad    M       = L_-  L_+, 
	\end{equation}      
	\item[  Sol2:] 
	\begin{equation}
		\tilde{M}       = L_+ M L_-, \qquad      L  = L_-  L_+,   \qquad      \tilde{L}  = L_+  L_-. \label{sol2}
	\end{equation}     
\end{itemize}
These solutions occur when $M$ and $L$, in Sol1 and Sol2 respectively, can be factorised exactly in terms of the intertwining operators. In the general case, such factorisation may take the form of polynomials $P_L$ and $P_M$  such that $L_+ L_-=P_M(M)$ and $L_- L_+=P_L(L)$, respectively. This is particularly relevant because, in the generic setting, there are many possible choices for $L_+$ and $L_-$ as operators with distinct differential order, see for instance \cite{correa2,correa3,correa4, andrianov2003nonlinear,ioffe2004susy}. In the standard intertwiner approach one focuses usually on the solution of type Sol2 where $L_+$ and $L_-$ have the same differential order. The factorisation in terms of the intertwining operators implies that the two Hamiltonians $L$ and $\tilde{L}$ are quasi-isospectral to each other.

In the standard intertwiner approach one focuses usually on the solution of type Sol2. The factorisation in terms of the intertwining operators implies that the two Hamiltonians $L$ and $\tilde{L}$ are quasi-isospectral to each other. One easily verifies that the two eigenvalue equations
\begin{equation}
	  \tilde{L} \phi_n = \tilde{E}_n \phi_n, \qquad \text{and}   \qquad L \psi_n = E_n \psi_n ,
\end{equation}  
are consistently solved with for the eigenfunction and quasi-isospectral energy relations
\begin{equation}
        \phi_{n+1} = \frac{1}{\sqrt{E_n}} L_+ \psi_n, \qquad  \psi_{n-1} = \frac{1}{\sqrt{\tilde{E}_n }} L_- \phi_n, \qquad E_n = \tilde{E}_{n+1} .
\end{equation}  
The last equation makes it clear that we can not match the spectra isospectrally, but in a quasi-isospectral fashion.

A well-known realisation for the typical second order operators is used in the supersymmetric analogue of quantum mechanics \cite{Witten:1981nf,Cooper:1982dm,Witten,Cooper,bagchi2000super} or where they are often referred to as Darboux-transformations \cite{darboux,crum,matveevdarboux}.  In this case $L$ is a standard  Hamiltonian of second order that can be factorised easily into first order operators
\begin{equation}
	L = - \partial_x^2 - u(x) = L_- L_+, \qquad \text{where} \,\,\, L_\pm = \pm \partial_x + W(x) ,
\end{equation}  
when expressing the potential in the form $u(x)=W'(x)-W^2(x)$.
Then, the ground state $\psi_0$ of $L$ is also annihilated by $L_+$
\begin{equation}
	L \psi_0  =0, \qquad L_+ \psi_0  =0, \qquad \text{with} \,\,\,   \psi_0 = \exp\left[- \int^x W(s) ds \right]  .
\end{equation}  
The operator $\tilde{M} =  L_+ M L_-$ commutes by construction with $ \tilde{L}  $.

\subsection{ A reversed intertwiner operator approach}

\noindent By exchanging the roles of $L$ and $M$ we can follow the modified procedure:
\begin{itemize}
	\item[S1':] Find an operator $M_+$ with common ground state $\phi_0$ to $M$ 
	\begin{equation}
		M \phi_0      =0,  \qquad     	M_+ \phi_0      =0.    \label{S1d}
	\end{equation} 
	\item[S2':] Use $M_+$ as a left intertwining operator and find a new operator $\tilde{M}'$ so that
	\begin{equation}
		M_+ M       =\tilde{M}' M_+.  \label{S2d}
	\end{equation}  
	\item[S3':] Find a second right intertwining operator $M_-$ so that
	\begin{equation}
		   M  M_-  = M_- \tilde{M}'           .    \label{S3d}
	\end{equation}  
	\item[S4':] Identify a new operator $\tilde{L}'$ that commutes with $\tilde{M}'$
	\begin{equation}
		\left[ \tilde{L}' ,\tilde{M}' \right]=0.     \label{S4d}
	\end{equation} 
	 \item[S5':] Repeat the steps S1'-S4' by starting with $\tilde{M}'$ instead of $M$ to find a new operator $M''$.
\end{itemize}
 In this case the formal solutions to the above scheme are:
\indent     \begin{itemize}
	\item[Sol1:] 
	\begin{equation}
		\tilde{L}'       = M_+  M_-, \qquad    L'     = M_-  M_+,  \label{sol1d}
	\end{equation}      
	\item[Sol2:] 
	\begin{equation}
		\tilde{L}'       = M_+ L M_-, \qquad      M'  = M_-  M_+,   \qquad      \tilde{M}'  = M_+  M_-,   \label{sol2d}
	\end{equation}     
\end{itemize}
where $L'$ and $M'$ commute with $M$ and $L$, respectively. Using the same reasoning as in the previous section one can now argue that $\tilde{L}' $ and $ L' $ as well as $\tilde{M}' $ and $ M' $ are quasi-isospectral to each other. The same comment made after (\ref{sol2}) applies here, with $L$ and $M$ interchanged. 
	
\subsection{From KdV-M-operators to Hirota-Satsuma Lax operators}

We now carry out the modified version of the above procedure for a concrete integrable system. We commence with the time-independent Lax operators of the KdV equation
\begin{equation}
	L= - \partial_x^2 - u + \rho, \qquad M = 4 \partial_x^3 + 6 u \partial_x  + 3 u_x,  \label{LMorig}
\end{equation} 
where $u(x)$ satisfies the time-independent KdV equation so that (\ref{LMcom}) holds
\begin{equation}
		\left[ L ,M \right]=  6 u u_x + u_{xxx}      =0.   \label{stKdV}
\end{equation} 
The arbitrary constant $\rho$ can be adjusted to achieve a particular asymptotic behaviour of $u(x)$. Next we carry out step S1'. First we note that taking the left intertwining operator to be a generic first order operator, the ground state is easily identified as  
\begin{equation}
	M_+ = \partial_x - f(x), \quad \phi_0 =e^{\int^x f(s) ds} ,   \label{linA}
\end{equation} 
for any arbitrary function $f(x)$. Acting on this state with $M$ we find that $ \phi_0 $ is a mutual zero mode if and only if
\begin{equation}
	4 f_{xx}+6 f \left(2 f_x+u\right)+4 f^3+3 u_x=0 .     \label{auxequ}
\end{equation} 
Assuming that the constraint (\ref{auxequ}) holds, we can carry out step S2' and construct the operator $\tilde{M}'$ from the intertwining relation (\ref{S2d}). We find 
\begin{equation}
\tilde{M}' = 4 \partial_x^3 + (6 u + 12 f_x) \partial_x - 6 f \left( 2 f^2 + 3 u + 4 f_x   \right)   .   \label{Mtildegen}
\end{equation} 
In step S3' we solve now the intertwining relation (\ref{S3d}) for $M_-$. For this purpose we first integrate the static KdV-equation (\ref{stKdV}) and subsequently the resulting equation when multiplied by $u_x$. We obtain 
\begin{equation}
	\left(u_x  \right)^2= 2 \left( c_2  + c_1 u -u^3   \right),    \label{intconstcc}
\end{equation} 
where $c_1$, $c_2$ are the two integration constants.  Next we use (\ref{intconstcc}) to solve (\ref{auxequ}) further by expressing $f$ in terms of $u$. A particular set of solutions is easily obtained by splitting (\ref{auxequ}) into two equations
\begin{equation}
	12 f f_x  + 3 u_x =0, \qquad \text{and} \qquad 4 f^3 +6 f u + 4 f_{xx} =0 .  \label{fconst}
\end{equation} 
The first equation is easily integrated out and when substituted into the second we obtain a constraint on the previously introduced integration constants 
\begin{equation}
	f(x) =\frac{1}{\sqrt{2}} \sqrt{\alpha - u}, \qquad \text{and} \qquad c_2 +  c_1 \alpha -\alpha^3  =0 . \label{fsol}
\end{equation} 
Here $\alpha$ is a new integration constant and $c_1$, $c_2$  are the constants introduced in (\ref{intconstcc}). With the help of these relations we can now solve for $M_-$ in terms of $u$ and its derivatives
\begin{equation}
	M_- = \partial_x^3 +  f \partial_x^2 + \frac{3}{2}    \left(u - \frac{u_x}{2 f} \right) \partial_x + \frac{3}{4} u_x - 3f(f^2+u),  \label{228}
\end{equation} 
where $f(x)$ is of the form in (\ref{fsol}). As we pointed out in Section 2, it is worth emphasising that the differential orders of the intertwining operators $M_+$ and $M_-$ are different. Although typically we find order-three operators such (\ref{228}) for the case at hand, in the examples below we will see that a second-order operator can be found when $u$ is taken to be a rational solution of the KdV equation.

Step S4' is now easily carried out. Given the fourth order operator of the form  
\begin{equation}
    \tilde{L}' =M_+ M_- ,     \label{mpmm}
\end{equation} 
it commutes by construction with $\tilde{M}'$ given the intertwining relations. We note that there is no second order operator $\tilde{L}' $ that commutes with $\tilde{M}'$. However, the fourth order operator (\ref{mpmm}) can be factorised into a product of two second order operators 
\begin{equation}
\tilde{L}' = H_1 H_2=  \left(  - \partial_x^2 + V_1 \right)  \left(  - \partial_x^2 + V_2 \right),
\end{equation} 
where the potentials are
\begin{equation}
 V_1 = \frac{\alpha}{2} - \frac{u}{2} + \frac{u_x}{2 \sqrt{2} \sqrt{\alpha -u}}, \qquad \text{and} \qquad
  V_2 = - \frac{3}{2}  u + \frac{ 3u_x}{2 \sqrt{2} \sqrt{\alpha -u}}.
\end{equation} 
In this form it remains unclear which system these operators might correspond to, but introducing the fields
\begin{equation}
    	\psi = \frac{\alpha}{4} - u + \frac{u_x}{\sqrt{2} \sqrt{\alpha -u}},  \qquad \text{and} \qquad
    		\phi = \frac{\alpha}{4} + \frac{u}{2} - \frac{u_x}{2 \sqrt{2} \sqrt{\alpha -u}},
\end{equation} 
the potentials become $V_1 = \phi +\psi$, $V_2 = \psi -\phi$. In turn, $\tilde{M}'$ and $\tilde{L}' $ acquire the forms 
\begin{eqnarray}
   \tilde{M}'&=&  4 \partial_x^3  \partial_x  +\left( \frac{3}{2} \alpha - 6 \psi \right) \partial_x  - 3 \psi_x + 6 \phi_x  ,  \\
   \tilde{L}' &=&  \partial_x^4 - 2 \psi \partial_x^2 - 2 (\psi_x -\phi_x) \partial_x -\psi_{xx} + \phi_{xx} + \psi^2 - \phi^2 .
\end{eqnarray} 
These two operators commute up to the equations of motion
\begin{eqnarray}
   \psi_{xxx} + \frac{3}{2}  \left( 8 \phi \phi_x - 4 \psi \psi_x + \alpha \psi_x      \right)     &=& 0 ,  \\
	\phi_{xxx}  - \frac{3}{4}  \left( \alpha + 4 \psi    \right)  \phi_x      &=& 0 ,
\end{eqnarray} 
which we identify as the coupled three field Hirota-Satsuma system \cite{hirota1981soliton} with two of the fields being identical. Given (\ref{sol1d}), we also verify that $L'$ commutes indeed with $M$. In fact it is a product of two standard $L$-operators. Stating the explicit $\rho$ dependence in (\ref{LMorig}), we have
\begin{equation}
	 L' =M_- M_+ = L(\rho_-)  L(\rho_+), \qquad \text{where} \quad \rho_\pm = \frac{1}{4}  \left(  \alpha \pm \sqrt{ 4 c_1 - 3 \alpha^2    }   \right)  .
\end{equation}	
The factorisation properties of $ L'$ and  $\tilde{L}'$ in terms of the left and right $M$-intertwining operators implies that they are quasi-isospectral. 

\subsection{Explicit solutions for the  KdV-$M$-operators}

Next we consider concrete solutions for the function $u(x)$ and also return to step S5' attempting to consecutively construct more $M$-operators. We specify some of the solutions $u$ in a systematic fashion. In view of equation (\ref{intconstcc}) we define the third order polynomial function in $u$
\begin{equation}
 P(u)= \frac{2}{\lambda} \left( c_2  + c_1 u -u^3   \right)  ,   \label{pol}
\end{equation} 
involving a constant $\lambda$. Separating variables a solution $u$ is then obtained from  (\ref{intconstcc}) by solving 
\begin{equation}
	\pm \sqrt{\lambda} (x-x_0) = \int du \frac{1}{\sqrt{P(u)}}, \label{solPu}
\end{equation} 
for $u$ after the integration of the right hand side has been carried out. Making particular assumptions about $P(u)$ the integral on the right hand side can be computed in a controlled manner and we obtain the well-known one-soliton solutions of the KdV equation in terms of rational, hyperbolic and elliptic functions. 

To achieve that we make some concrete assumptions on $P(u)$ to solve (\ref{solPu}) for $u$ and hence (\ref{fconst}), or more generally (\ref{auxequ}), for $f$.

\subsubsection{The rational function solution, infinite quasi-isospectral sequences} 
Assuming that $P(u) = (u-A)^3$ for some constant $A$, the solution resulting from (\ref{solPu}) is 
\begin{equation}
	u(x) = A+\frac{4}{\lambda  \left(x-x_0\right)^2}  .
\end{equation} 
The only possibility to match $P(u)$ with the polynomial in (\ref{pol}) is to take $\lambda=-2$ and $A=c_1=c_2=0$, so that 
\begin{equation}
	u(x) = -\frac{2}{ \left(x-x_0\right)^2}  .
\end{equation} 
Consequently equation (\ref{fsol}) yields $\alpha=0$ and therefore $f(x)=1/(x-x_0)$. Then the new Hamiltonian (\ref{Mtildegen}) acquires the form 
\begin{equation}
	\tilde{M} '= 4    \left[ \partial_x^3 - \frac{6}{ (x-x_0)^2 }    \partial_x + \frac{12}{ (x-x_0)^3 }   \right] .
\end{equation} 
Notice that $\psi_0=x-x_0$, resulting from (\ref{linA}), is not an eigenstate of $L$, but is a Jordan state satisfying $L^2 \psi_0 =0$ for $\rho=0$. The zero modes of $M$ are not unique. Further solutions are $\psi_0^1=(x-x_0)^{-1}$ and $\psi_0^3=(x-x_0)^{3}$, which are also zero modes for $L$ and $L^3$, respectively. Using these states in S1' of the construction procedure produces different intertwining operators and therefore also different $\tilde{M}' $-operators. This process may be continued to the next level and in principle ad infinitum. Denoting $M= M_0$, $\tilde{M}= M_1$, etc.,  we have in general the following expressions when iterating from 
\begin{equation}
	M_n =  4     \partial_x^3 + \frac{a_n}{ (x-x_0)^2 }    \partial_x + \frac{b_n}{ (x-x_0)^3 }   ,   \label{Mopgen}
\end{equation} 
to the next level
\begin{equation}
	M_{n+1} =  4     \partial_x^3 + \frac{a_{n+1}}{ (x-x_0)^2 }    \partial_x + \frac{b_{n+1}}{ (x-x_0)^3 }   , \label{Mopgen1}
\end{equation} 
with coefficients $a_n,b_n,a_{n+1},b_{n+1} \in \mathbb{Z}$ and $n \in \mathbb{N}_0$. Assuming the zero mode of $M_n $ to be of the form $\psi_0^n =(x-x_0)^k$, the intertwining operator results to $M_+^n = \partial_x - f_n(x)$ where $f_n = [\ln(\psi_0^n)]_x = k/(x-x_0)$. The intertwining relation 
\begin{equation}
	M_+^n M_n = M_{n+1}  M_+^n, \qquad n \in \mathbb{N}_0,     \label{intrel2}
\end{equation} 
holds when 
\begin{eqnarray}
	0&=& k a_n + b_n +4k(k-1)(k-2),   \label{abn} \\
	a_{n+1}&=& a_n -12 k, \label{abn1}  \\
	b_{n+1}&=& b_n - 2 a_n -12 k(k-2) . \label{abn2} 
\end{eqnarray} 
where the coefficients $(a_n,b_n)$, $(a_{n+1},b_{n+1})$ are introduced in (\ref{Mopgen}), (\ref{Mopgen1}). Thus, for given $(a_n,b_n)$ we may solve (\ref{abn}) for $k$ and compute  $(a_{n+1},b_{n+1})$ from (\ref{abn1}), (\ref{abn2}). Noting that (\ref{abn}) will always give three solutions for $k$ we generate in this way the sequence
\begin{equation}
	\!\! M_0 \xrightarrow{k_i} 	M_1^i \xrightarrow{k_j} 	M_2^{ij} \xrightarrow{k_k} 	M_3^{ijk}   \xrightarrow{k_\ell} 	M_4^{ijk\ell}  \xrightarrow{k_m} 	M_5^{ijk\ell m} \rightarrow \ldots  \quad i,j,k, \ell , m \in \{  1,2,3 \} .   \label{sequ}
\end{equation} 
Using the abbreviated notation $M_n(a_n,b_n)$ for the operator in (\ref{Mopgen}), we find
\begin{eqnarray*}
M_0(-12,12) :\,  &\xrightarrow{-1} & M_1^1(0,0), \,\, \xrightarrow{1}  M_1^2(-24,48) \,\,  \xrightarrow{3}  M_1^3(-48,0),  \\
M_1^1(0,0) :\  & \xrightarrow{0} & M_2^{11}=M_1^1, \,\,  \xrightarrow{1}  M_2^{12}=M_0 \,\,  \xrightarrow{2}  M_2^{13}(-24,0) :\,  \\
M_1^2(-24,48) :\,   &\xrightarrow{-2}  &M_2^{21}=M_1^1, \,\,  \xrightarrow{2}  M_2^{22}(-48,96), \,\,  \xrightarrow{3}  M_2^{23}(-60,60), \\
 M_1^3(-48,0) :\, &\xrightarrow{-2} &M_2^{31}=M_2^{13}   , \,\,  \xrightarrow{0}  M_2^{32}=M_2^{22}, \,\,  \xrightarrow{5}  M_2^{33}(-108,-84), \\
M_2^{13}(-24,0) :\,  &\xrightarrow{-1} & M_3^{131}= M_0 , \,\,  \xrightarrow{0}  M_3^{132}= M_1^2, \,\,  \xrightarrow{4}  M_3^{133}(-72,-48), \\
M_2^{22}(-48,96) :\,  &\xrightarrow{-3} & M_3^{221}= M_0 , \,\,  \xrightarrow{2}  M_3^{222}(-72,192) , \,\,  \xrightarrow{4}  M_3^{223}(-96,96), \\
M_2^{23}(-60,60) :\, &\xrightarrow{-3} & M_3^{231}=M_2^{13} , \,\,  \xrightarrow{1}  M_3^{232}= M_3^{222}, \,\,  \xrightarrow{5}  M_3^{233}(-120,0), \\
M_2^{33}(-108,-84) :\, &\xrightarrow{-3} & M_3^{331}=M_3^{133}  , \,\,  \xrightarrow{1}  M_3^{332}= M_3^{223}, \,\,  \xrightarrow{7}  M_3^{333}(-192,-288), \\
M_3^{133}(-72,-48) :\, &\xrightarrow{-2} & M_4^{1331}=M_1^3 , \,\,  \xrightarrow{-1}  M_4^{1332}= M_2^{23}, \,\,  \xrightarrow{6}  M_4^{1333}(-144,-192), \\
M_3^{222}(-72,192) :\, &\xrightarrow{-4} & M_4^{2221}=M_1^2 , \,\,  \xrightarrow{3}  M_4^{2222}(-108,300), \,\,  \xrightarrow{4}  M_4^{2223}(-120,240), \\
M_3^{223}(-96,96) :\,  &\xrightarrow{-4} & M_4^{2231}=M_1^3 , \,\,  \xrightarrow{1}  M_4^{2232}=M_4^{2222}, \,\,  \xrightarrow{6}  M_4^{2233}(-168,0), \\
M_3^{233}(-120,0) :\,  &\xrightarrow{-4} & M_4^{2331}=M_3^{133} , \,\,  \xrightarrow{0}  M_4^{2332}=M_4^{2223}, \,\,  \xrightarrow{7}  M_4^{2333}(-204,-180), \\
M_3^{333}(-192,-288) :\,  &\xrightarrow{-4} & M_4^{3331}=M_4^{1333} , \,\,  \xrightarrow{-2}  M_4^{3332}=M_4^{2233}, \,\,  \xrightarrow{9}  M_4^{3333}(-300,-660), \\
	&\vdots& .
\end{eqnarray*} 
As indicated in (\ref{sequ}), by constructing intertwining operators from various zero modes, we can generate a diverse array of sequences that bifurcate at each level. Some of the sequences are finite, meaning that the iteration will eventually return back to the original $M$-operator, such as for instance  
\begin{equation}
M_0  \xrightarrow{-1}  M_1^1  \xrightarrow{2} M_2^{13}  \xrightarrow{0}   M_2^{22}  \xrightarrow{-3}  M_0   .  
\end{equation} 
However, there are other sequences that are infinite and can even be expressed in a closed form. We identify 
\begin{eqnarray*}
	M_0  \xrightarrow[L]{-1}  M_1^1  \xrightarrow[L^2]{2} M_2^{13}  \xrightarrow[L^3]{4}   M_3^{133}  \xrightarrow[L^4]{6}  M_4^{1333} \ldots \!\!&\Rightarrow& M_n\left[ -12(n-1)n , -8 (2n-3n^2 + n^3   )\right],    \\
		M_0  \xrightarrow[L^2]{1}  M_1^2  \xrightarrow[L^3]{3} M_2^{23}  \xrightarrow[L^4]{5}   M_3^{233}  \xrightarrow[L^5]{7}  M_4^{2333} \ldots \!\!&\Rightarrow& M_n\left[ -12(1+n^2) , 4 (3+8n+3n^2-2n^3)\right],   \\
		M_0  \xrightarrow[L^3]{3}  M_1^3  \xrightarrow[L^4]{5} M_2^{33}  \xrightarrow[L^5]{7}   M_3^{333}  \xrightarrow[L^6]{9}  M_4^{3333} \ldots \!\!&\Rightarrow& M_n\left[ -12(1+n)^2 , 4 (3+2n-3n^2-2n^3)\right]  ,
\end{eqnarray*} 
for $n=1,2,\ldots$. At each level we have to decide which particular zero mode of the possible three we are taking to continue the sequence. Our choice is guides by selecting the states as being Jordan zero modes of increasing powers of $L$. Below the arrow we have written the operator for this is the case, i.e. whose zero mode is also a zero mode of $M_n$. The states are easily identified by using the fact that $L^n(\rho=0) (x-x_0)^k = 0$ for $k=-1,1,2, \ldots, 2n-3,2n-2,2n$. Generating these sequences leads to Hamiltonians that are all  isospectral to each other, up to the ground state that we loose each time we proceed to the next level.

In summary, we have the three infinite sequences as solution to the intertwining relation (\ref{intrel2}) in step S2'
\begin{eqnarray}
		M_n &=&  4  \left[    \partial_x^3 - \frac{ 3(n-1)n }{ (x-x_0)^2 }    \partial_x - \frac{2 (2n-3n^2 + n^3   )}{ (x-x_0)^3 }  \right],  \qquad 	M_+^n = \partial_x - \frac{2n}{(x-x_0)},  \label{MMp1} \\
			M_n &=&  4  \left[   \partial_x^3 - \frac{ 3(1+n^2) }{ (x-x_0)^2 }    \partial_x + \frac{ 3+8n+3n^2-2n^3}{ (x-x_0)^3 }  \right],  \qquad 	M_+^n = \partial_x  - \frac{1+2n}{(x-x_0)},   \label{MMp2} \\
				M_n &=&  4  \left[    \partial_x^3 - \frac{ 3(1+n)^2}{ (x-x_0)^2 }    \partial_x + \frac{3+2n-3n^2-2n^3}{ (x-x_0)^3 }   \right],  \qquad 	M_+^n = \partial_x  - \frac{3+2n}{(x-x_0)}.  \qquad  \label{MMp3}
\end{eqnarray} 
Next we try to find $M_-^n$ from the intertwining relation (\ref{S4d}) in S4'.  Insisting on third order operators, it turns out that this can only be achieved for specific values of $n$, but not the entire infinite series. However, we will see that a second order operators are valid solutions for the entire infinite series. We find
the following particular solutions together with the corresponding $\tilde{L}'$ when we consider third order operators as $M_-^{n}$. We use the same
conventions to introduce graded $L$-operators. For the series (\ref{MMp1}) we find
\begin{eqnarray}
	M_-^{- 1}&=& \partial_x^3 - 2 f  \partial_x^2 -  u  \partial_x, \qquad L_{- 1} = M_+^{-1}  M_-^{- 1}=  \partial_x^4 , \\
       M_-^0 &=&  \partial_x^3,   \qquad \qquad \qquad \qquad \,\,\,\,   L_{0} = M_+^{0}  M_-^{0} = \partial_x^4 , \\
        M_-^{1} &=& \partial_x^3 + 2 f  \partial_x^2 +  u  \partial_x, \qquad\,\,\,\, L_{ 1} = M_+^{1}  M_-^{1}=  \partial_x^4 + 4 u \partial_x^2 + 2 u_x \partial_x .
\end{eqnarray} 
 For the series (\ref{MMp2}) we find the solutions
 \begin{eqnarray}
 	M_-^{- 2}&=& \partial_x^3 - 3 f  \partial_x^2 + u  \partial_x +\frac{5}{2} u_x , 
 	   \quad L_{- 2} = M_+^{-2}  M_-^{- 2}=  \partial_x^4  + 4 u \partial_x^2  + 2 u_x \partial_x     , \\
 	M_-^{- 1}&=& \partial_x^3 -  f  \partial_x^2 + 2 u  \partial_x + u_x, 
 	\quad \,\, \, \,\,L_{- 1} = M_+^{-1}  M_-^{- 1}=  \partial_x^4+ 2 u \partial_x^2  + 2 u_x \partial_x  + \frac{2}{3} u_{xx} , \qquad \,\,\,\,  \\
 	M_-^0 &=&  \partial_x^3 + f  \partial_x^2 + 3 u  \partial_x +\frac{3}{2} u_x,   
 	\quad \,\,\,\,\,   L_{0} = M_+^{0}  M_-^{0} = \partial_x^4+ 4 u \partial_x^2  + 6 u_x \partial_x + 2 u_{xx} , \\
 	M_-^{1} &=& \partial_x^3 + 3 f  \partial_x^2 +  4 u  \partial_x, 
   \qquad	\qquad\,\,  L_{ 1} = M_+^{1}  M_-^{1}=  \partial_x^4 + 10 u \partial_x^2 + 10 u_x \partial_x   ,
 \end{eqnarray} 
 and for the series (\ref{MMp3}) the only solutions are 
  \begin{eqnarray}
 	M_-^{- 2}&=& \partial_x^3 -  f  \partial_x^2  , 
 	\quad L_{- 2} = M_+^{-2}  M_-^{- 2}=  \partial_x^4   , \\
 	M_-^{- 1}&=& \partial_x^3 +  f  \partial_x^2 +  u  \partial_x + \frac{1}{2}u_x, 
 	\quad \,\, \, \,\,L_{- 1} = M_+^{-1}  M_-^{- 1}=  \partial_x^4+ 2 u \partial_x^2  + 2 u_x \partial_x  + \frac{2}{3} u_{xx} . \qquad \,\,\,\, 
 \end{eqnarray} 
 
 Given the breakdown of the factorization property in terms of a third-order intertwining operator $M_-^{n}$ along the entire series, one can naturally address two related questions simultaneously: (i) whether a generic operator 
 $M_-^{n}$ can be defined for a given order, and (ii) whether it is possible to construct shape-invariant differential operators that factorize consistently for the entire infinite series. Here we extend the notion of shape
 invariant potentials, see e.g. \cite{Levai:1989eaa}, to differential operators, that is we consider operators of
 the same functional form with different values for the constants involved. The answer is
 positive in both cases: We define the operator
\begin{eqnarray} \label{h1}
	M_{n,m,\mu}&=&4 \left[ \partial_x^3 -\frac{\mu ^2+m^2+m+3 n^2-3 n+\mu  (m+3 n-1)}{x^2} \partial_x  \right.    \qquad \qquad  \qquad  \\
	&& \left. + \frac{(m-n+2) (\mu +2 n) (\mu +m+n-1)}{x^3} \right] , \notag
\end{eqnarray}
depending on three parameters $n, m$ and $\mu$,  that we interpret as a third order Hamiltonian. Evidently $M_{n,m,\mu}$ is of the same functional form as the operators $M_n$ in (\ref{MMp1})-(\ref{MMp3}) with $x_0=0$ for simplicity. It turns out this operator can be factorised into left and right intertwining operators 
\begin{equation}
	   	M_{n,m,\mu} = 4 M^-_{n,m,\mu} M^+_{n,m,\mu} ,
\end{equation} 
with
\begin{equation}
	M^-_{n,m,\mu}= \partial_x^2  + \frac{2n + \mu}{x} \partial_x  + \frac{(n-m-1)(m+n+\mu)}{x^2}, \qquad             M^+_{n,m,\mu} = \partial_x - \frac{2n+ \mu}{x},
\end{equation} 
satisfying the intertwining relations (\ref{S2d}) and (\ref{S3d}) in the form
\begin{equation}
    M^+_{n,m,\mu}  	M_{n,m,\mu}  = 	M_{n+1,m,\mu} M^+_{n,m,\mu} , \qquad
     	M_{n,m,\mu}   M^-_{n,m,\mu}  =  M^-_{n,m,\mu}  M_{n+1,m,\mu}  .
\end{equation} 
The previous cases (\ref{MMp1}), (\ref{MMp2}) and (\ref{MMp3}) are recovered for $(m,\mu) = \left\{ (0,0), (1,1), (-1,3)    \right\}$.

\subsubsection{The hyperbolic function solution} 
Assuming next that $P(u) = (u-A)^2(u-B)$ for some constants $A$ and $B$, the solution resulting from (\ref{solPu}) is 
\begin{equation}
	u(x) = B+(A-B) \tanh ^2\left[  \frac{1}{2} \sqrt{\lambda } \sqrt{A-B} (x-x_0) \right]  . \label{solhyp}
\end{equation} 
Matching $P(u)$ with the polynomial in (\ref{pol}) we obtain $\lambda=-2$,  $A= -\kappa$, $B= 2 \kappa$, $c_1= 3 \kappa^2$, $c_2= 2 \kappa^3$, where the constant $\kappa$ is related to the asymptotic value of $u$ as $-\kappa := \lim_{ \vert x \vert \rightarrow\infty} u$. With the constants identified as specified the solution (\ref{solhyp}) reduces to
\begin{equation}
	u(x) =2 \kappa -3 \kappa  \tanh^2\left[ \sqrt{\frac{3}{2}} \sqrt{\kappa } (x-x_0)\right] .
\end{equation} 
Here we wish to focus on the hyperbolic case and therefore make the convenient choice $\kappa = 2/3$ for the asymptotic value. For simplicity we also set $x_0=0$ in what follows, but evidently this constant may be re-introduce by a simple shift $x\rightarrow x-x_0$. In this case we obtain 
\begin{equation}
	u(x) = -\frac{2}{3}  + 2 \sech^2 x .   \label{hypsolsimp}
\end{equation} 
The auxiliary functions $f$ are now computed to
\begin{equation}
	f_s =  2 \csch( 2 x ),  \quad  f_b(x) = - \tanh x,   \quad 
	f_n(x) = \tanh x  ,  \label{eigenstf}
\end{equation} 
where $f_s$ is a direct solution of (\ref{auxequ}) and $f_b,f_n$ are computed from the stronger constraint (\ref{fsol}). The corresponding zero modes (\ref{linA}) of $M$ are therefore
\begin{equation}
	\psi^0_s(x) = \tanh x ,  \quad  \psi^0_b(x) = \sech x,   \quad 
	\psi^0_n(x) = \cosh x  .   \label{eigenst}
\end{equation} 
Interestingly, $\psi_s^0$ is one of the scattering states $\psi_k^\pm(x) = \left(  \tanh x \mp i k   \right) e^{\pm i k x}$ of the reflectionless potential, $ \psi^0_b(x) $ is the zero mode bound state of the $L$-operator for $\rho = 1/3$ and $\psi^0_n$ is a Jordan state for the same value of $\rho$, i.e.
\begin{equation}
	L \psi_k^\pm(x)  = (1+k^2)  \psi_k^\pm(x)   ,   \qquad   L \psi^0_b(x)  = 0,  \qquad 
	L^2 \psi^0_n=0.
\end{equation}
Acting with the asymptotic expression of the operator $M_\infty := \lim_{\vert x\vert \rightarrow \infty } M =4 \partial_x^3 - 4 \partial_x $ on the asymptotic state  $\lim_{\vert x\vert \rightarrow \infty } \psi_k^\pm(x)  \sim (1\mp ik) e^{\pm i k x }$ it is then easily verified that
\begin{equation}
	M \psi_k^\pm(x)  =\mp 4 i ( k+k^3)  \psi_k^\pm(x)  .
\end{equation}
The left intertwining operators $M^+_i$ for $i=s,b,n$ are trivially obtained from (\ref{linA}) with $f(x)$ taken in the form (\ref{eigenst}), and from (\ref{Mtildegen}) the three quasi-isospectral  Hamiltonians are computed to 
\begin{eqnarray}
	\tilde{M}'_s &=& 4 \partial_x^3 - 4 \left(1+ 3 \text{csch}^2 x \right) \partial_x +12 \coth x \, \text{csch}^2 x  , \\ 
	\tilde{M}'_b &=& 4 \partial_x^3 - 4  \partial_x  = M_\infty ,\\
	\tilde{M}'_n &=& 4 \partial_x^3 - 4 \left(1 - 6  \sech^2 x \right) \partial_x  - 48 \tanh x  \sech^2 x  . 
\end{eqnarray}
The right intertwining operators are obtained as
\begin{eqnarray}
	{M}^-_s &=& \partial_x^3+2\csch (2x) \partial_x^2-2 \csch^2 x  \partial_x+2 \tanh  x
	\left(\coth ^4 x -1\right)  , \\ 
	{M}^-_b &=& \partial_x^3 -\tanh x \partial_x^2  - \partial_x + \tanh x  ,\\
	{M}^-_n &=&  \partial_x^3 +\tanh x \partial_x^2+\left(6 \sech^2 x -1\right) \partial_x-\left[\tanh x
	\left(6 \text{sech}^2 x +1\right)\right] . 
\end{eqnarray}
Here ${M}^-_b $ and ${M}^-_n$ are directly obtained from (\ref{Mtildegen}), whereas ${M}^-_s $ needed to be computed from scratch as $f_s$ does not satisfy (\ref{fsol}).
We verify that the factorised operator $\tilde{L}'_i=  {M}^+_i {M}^-_i$ commutes with $\tilde{M}'_i $, for $i=s,b,n$. By construction the new operators $\tilde{L}''_i=  {M}^-_i {M}^+_i$ are quasi-isospectral to $\tilde{L}'_i$.

\subsubsection{The Jacobi elliptic function solution} 
Proceeding as in the previous sections, we solve now (\ref{solPu}) with the assumption $P(u) = (u-A)(u-B)(u-C)$, with constants $A$, $B$ and $C$. This ansatz leads to a general solution of the form
\begin{equation}
	u(x)= \frac{2}{3} (m-2)c^2 +2 c^2  \text{dn}^2\, ( c x \vert m) ,
\end{equation}
where $c$ is a free parameter and $0\leq m \leq 1$ denotes the elliptic modulus. A convenient choice to fix the free parameter $c$ is to demand that we recover the previous solution in the hyperbolic limit. Thus, we take from now one $c=1$ and note that with $\lim_{m \rightarrow 1} \text{dn}\, (x \vert m) =\sech x $ we recover precisely the solution in (\ref{hypsolsimp}) for $m\rightarrow 1$.
The auxiliary functions $f$ are now computed to
\begin{equation}
	f_s =  \frac{\text{cn}\, (x\vert m)\text{dn}\, (x\vert m)}{\text{sn}\, (x\vert m)},  \quad  f_b(x) =-\frac{\text{dn}\, (x\vert m)\text{sn}\, (x\vert m)}{\text{cn}\, (x \vert m)},   \quad 
	f_n(x) = \sqrt{m} \, \text{sn}\, (x \vert m) ,  \label{eigenstell}
\end{equation} 
where all $f_i$  for $i=s,b,n$ are solutions of (\ref{auxequ}), but only $f_s$ satisfies the stronger constraint (\ref{fsol}). The zero modes of $M$ are directly calculated from (\ref{linA}) to
\begin{equation}
	\phi^0_s(x) =  \text{sn}\, (x\vert m)  ,  \quad  \phi^0_b(x) = \text{cn}\, (x\vert m) ,   \quad 
	\phi^0_n(x) = \frac{1}{\sqrt{1-m}}   \left[ \text{dn}\, (x\vert m)  -\sqrt{m} \text{cn}\, (x\vert m)     \right]                  \label{eigenstellell}
\end{equation} 
In the limit $m \rightarrow 1$, we recover the expression for $f_i$ in (\ref{eigenstf}) from the hyperbolic solutions in a straightforward manner. Also $\lim_{m \rightarrow 1} \phi^0_s =\psi^0_s $ and $\lim_{m \rightarrow 1}\phi^0_b =  \psi^0_b $ are computed in a straightforward manner.  The limit $\lim_{m \rightarrow 1} \phi^0_n $ is more subtle and can not be computed directly from the expression in (\ref{eigenstellell}). The reason is that in the integrals used in (\ref{linA})  the integration and the limit do not commute, i.e. $\lim_{m \rightarrow 1} \left( \int^x f_n dx \right) \neq  \int^x  \left(\lim_{m \rightarrow 1}f_n  \right) dx $. The left intertwining operators $M^+_i$ for  $i=s,b,n$ are trivially obtained from (\ref{linA}) with $f(x)$ taken in the form (\ref{eigenstell}), and from (\ref{Mtildegen}) the three isospectral  Hamiltonians are computed to 
\begin{eqnarray}
	\tilde{M}'_s &=& 4 \partial_x^3 + 4 \left(1+m-\frac{3}{\text{sn}(x|m )^2}\right)\partial_x +12\frac{\text{cn}(x|m ) \text{dn}(x|m )}{\text{sn}(x|m )^3}  , \\ 
	\tilde{M}'_b &=& 4 \partial_x^3 + 4 \left( 1-2 m + 3  \frac{ m -1}{\text{cn}(x|m )^2} \right) \partial_x + 12 (m -1 )\frac{\text{dn}(x|m ) \text{sn}(x|m )}{\text{cn}(x|m )^3} ,\\
	\tilde{M}'_n &=& 4 \partial_x^3 +4 \left(3 \text{dn}(x|m) \left(\sqrt{m} \text{cn}(x|m)+\text{dn}(x|m)\right)+m-2\right) \partial_x \\
	&& -12 \sqrt{m} \text{sn}(x|m) \left(2 \text{dn}(x|m) \left(\sqrt{m}
	\text{cn}(x|m)+\text{dn}(x|m)\right)+m-1\right) .
\end{eqnarray}

Since the $f_n$-function satisfies (\ref{fsol}), we directly apply (\ref{Mtildegen}) to compute the corresponding  right intertwining operators $M^-_i$. For the other two cases we obtain the operators by direct computation  
\begin{eqnarray}
	{M}^-_s &=& \partial_x^3+f_s \partial_x^2-\frac{2}{\text{sn}(x|m )^2}  \partial_x+\frac{2 \text{cn}(x|m ) \text{dn}(x|m )}{\text{sn}(x|m )^3} , \\ 
	{M}^-_b &=& \partial_x^3 + f_b\partial_x^2 
	+ \frac{2 m -2}{\text{cn}(x|m )^2} \partial_x + \frac{2 (m -1) \text{dn}(x|m ) \text{sn}(x|m )}{\text{cn}(x|m )^3} ,\\
	{M}^-_n &=&    \partial_x^3+f_n \partial_x^2 +
	  \left[ 3 \text{dn}(x|m) \left(\sqrt{m} \text{cn}(x|m)+\text{dn}(x|m)\right)+m-2 \right] \partial_x \\
	     && -\sqrt{m} \text{sn}(x|m) \left[ 3 \text{dn}(x|m) \left(\sqrt{m}  \text{cn}(x|m)+\text{dn}(x|m)\right)+2
	     m-1\right] . \quad \notag
\end{eqnarray}
We verify that the factorised operator $\tilde{L}'_i=  {M}^+_i {M}^-_i$ commutes with $\tilde{M}'_i $, for $i=s,b,s$. Once more, by construction the new operators $\tilde{L}''_i=  {M}^-_i {M}^+_i$ are quasi-isospectral to $\tilde{L}'_i$. All expressions in this section give the same expressions from the previous section in the limit $m \rightarrow 1$.

\section{Conclusions}

We have introduced a reversed Lax pair construction as a novel method for generating quasi-isospectral higher-order Hamiltonians. By designating the higher-order 
$M$-operator as the starting point, we systematically constructed new Hamiltonians through intertwining techniques, demonstrating that these operators are quasi-isospectral to each other with spectra differing by at least one eigenstate. 

This reversed perspective provides a new structural understanding of integrable systems and opens up a new class of exactly solvable models characterised by higher-order dynamics. The methodology was exemplified through explicit constructions based on the time-independent KdV system and its extensions, revealing novel quasi-isospectral partners for the rational, hyperbolic, and elliptic solutions. Particularly striking is the demonstration of infinite sequences of such Hamiltonians, which were shown to exhibit shape-invariant structures and satisfy consistency conditions via higher-order intertwining relations.

Our analysis supports the broader utility of higher-order conserved charges, usually overlooked as candidates for Hamiltonians, in capturing rich spectral phenomena including degeneracies, spectral singularities, and resonances.

With regards to future extensions, this framework paves the way for several promising research directions. One avenue includes exploring its extension to non-Hermitian/${\cal PT}$-symmetric quantum systems, where quasi-isospectrality may interact in subtle ways with complex and even real spectral structures. In a reversed space-time setting these newly proposed systems may serve a candidates for higher time-derivative theories as indicated in the introduction. We have explored the first non-trivial example of a third-order Lax operator $M$, built upon the KdV hierarchy and relating it to a particular case of the Hirota-Satsuma one. A natural next step is to consider higher-order Lax operators and/or their higher-order intertwining operators as initial step. Clearly, this idea extends beyond the KdV hierarchy and can also be applied to other nonlinear hierarchies, such as the reductions of the Ablowitz–Kaup–Newell–Segur (AKNS) hierarchy. These insights might reveal connections with established and yet unexplored integrable hierarchies.

In summary, the reversed Lax pair approach offers a fertile ground for further discoveries in the theory of integrable systems and offers new possibilities for constructing exactly solvable models in quantum mechanics and field theory.

 \medskip

\noindent {\bf Acknowledgments}: FC was supported by Fondecyt Grants No. 1211356 and No. 1252036. FC acknowledges the warm hospitality and support of the Department of Mathematics, City St George’s, University of London.

\newif\ifabfull\abfulltrue

\end{document}